\documentclass[iop]{emulateapj}

\usepackage{natbib}
\usepackage{graphicx}
\usepackage{multirow}

\def\sun{\ifmmode\odot\else$\odot$\fi}

\shorttitle{Deeply embedded AGN 
activity in the nuclear regions of Arp~299}
\shortauthors{Alonso-Herrero  et al.}

\begin{document}

\title{Uncovering the deeply embedded AGN activity in the nuclear regions of the
  interacting galaxy Arp~299}
\author{A. Alonso-Herrero\altaffilmark{1,2}, 
  P. F. Roche\altaffilmark{3}, 
  P. Esquej\altaffilmark{4}, 
  O. Gonz\'alez-Mart\'{\i}n\altaffilmark{5,6}, 
  M. Pereira-Santaella\altaffilmark{7},
  C. Ramos Almeida\altaffilmark{5,6}, 
  N. A. Levenson\altaffilmark{8},
  C. Packham\altaffilmark{9},
  A. Asensio Ramos\altaffilmark{5,6},
  R. E. Mason\altaffilmark{10},
  J. M. Rodr\'{\i}guez Espinosa\altaffilmark{5,6},
  C. Alvarez\altaffilmark{5,6},
  L. Colina\altaffilmark{4},  
  I. Aretxaga\altaffilmark{11},  
  T. D\'{\i}az-Santos\altaffilmark{12},  
  E. Perlman\altaffilmark{13},  and
  C. M. Telesco\altaffilmark{14}  
} 
\altaffiltext{1}{Instituto de F\'{\i}sica de
  Cantabria, CSIC-UC, 39005 Santander, Spain;
  E-mail: aalonso@ifca.unican.es} 
\altaffiltext{2}{Augusto Gonz\'alez Linares Senior Research Fellow}
\altaffiltext{3}{Astrophysics Department, University
of Oxford, Oxford OX1 3RH, UK}
\altaffiltext{4}{Centro de Astrobiolog\'{\i}a, CSIC-INTA, 28035
  Madrid, Spain}
\altaffiltext{5}{Instituto de Astrof\'{\i}sica de Canarias, 38205 La
  Laguna, Spain}
\altaffiltext{6}{Universidad de la Laguna, 38205 La
  Laguna, Spain}
\altaffiltext{7}{Istituto di Astrofisica e Planetologia Spaziali,
  INAF, 00133 Rome, Italy} 
\altaffiltext{8}{Gemini Observatory, La Serena, Chile}
\altaffiltext{9}{University of Texas at San Antonio, San Antonio, TX 78249}
\altaffiltext{10}{Gemini Observatory, Hilo HI 96720}
\altaffiltext{11}{INAOE, 72000 Puebla, Mexico}
\altaffiltext{12}{Spitzer Science Center, Caltech, Pasadena, CA 91125}
\altaffiltext{13}{Florida Institute of Technology, Melbourne, FL 32901}
\altaffiltext{14}{Department of Astronomy, University of Florida,
  Gainesville, FL 32611} 

\begin{abstract}
We present mid-infrared (MIR) $8-13\,\mu$m spectroscopy of the
nuclear regions of the interacting galaxy Arp~299 (IC~694+NGC~3690)
obtained with 
CanariCam (CC) on the 10.4\,m Gran Telescopio Canarias (GTC). 
The high angular resolution ($\sim 0.3-0.6\arcsec$) of the
data allows us to probe nuclear physical scales between 60 and 
120\,pc, which is a factor of 10 improvement over previous MIR spectroscopic
observations of this system. The GTC/CC spectroscopy displays evidence
of 
deeply embedded Active Galactic Nucleus (AGN) activity in both
nuclei. The GTC/CC 
nuclear spectrum of NGC~3690/Arp~299-B1 can be explained as emission 
from AGN-heated dust  in a clumpy torus with both a high covering 
factor and high extinction along the line of sight. The estimated
bolometric 
 luminosity of the AGN in NGC~3690 is $3.2 \pm 0.6 \times 
10^{44}\,{\rm erg \,s}^{-1}$. The nuclear GTC/CC spectrum of
IC~694/Arp~299-A 
shows $11.3\,\mu$m polycyclic aromatic hydrocarbon (PAH) emission 
stemming from a deeply embedded ($A_V \sim 24\,$mag)
region of less than 120\,pc in size. There is also  a
continuum-emitting dust component. If associated with the
putative AGN in IC~694, we  estimate that it would be approximately 5
times less  luminous than the AGN in NGC~3690. The presence of dual
AGN activity makes Arp~299 a good example to study such phenomenon in the early
coalescence phase of interacting galaxies. 
\end{abstract}
\keywords{galaxies: nuclei --- galaxies: Seyfert ---
  infrared: galaxies --- galaxies: individual (Arp~299, NGC~3690, IC~694)}

\section{Introduction}\label{s:intro}
\begin{figure*}

\hspace{2.5cm}
\includegraphics[width=0.3\textwidth,angle=-90]{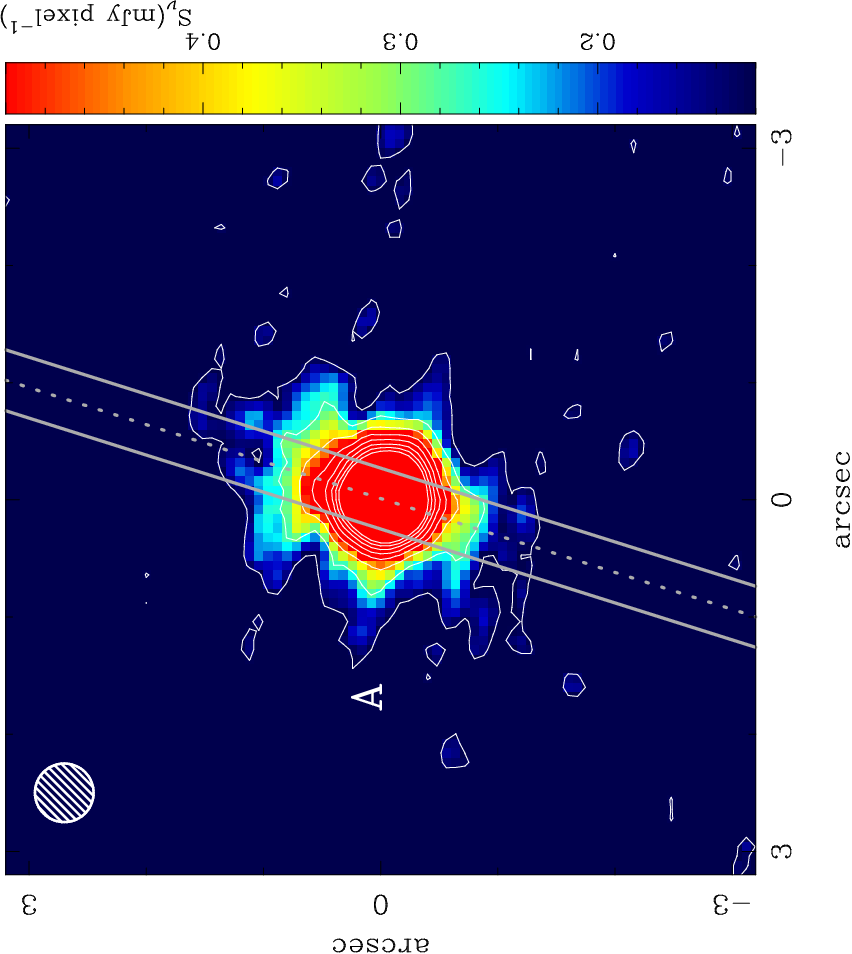}
\hspace{0.5cm}
\includegraphics[width=0.3\textwidth,angle=-90]{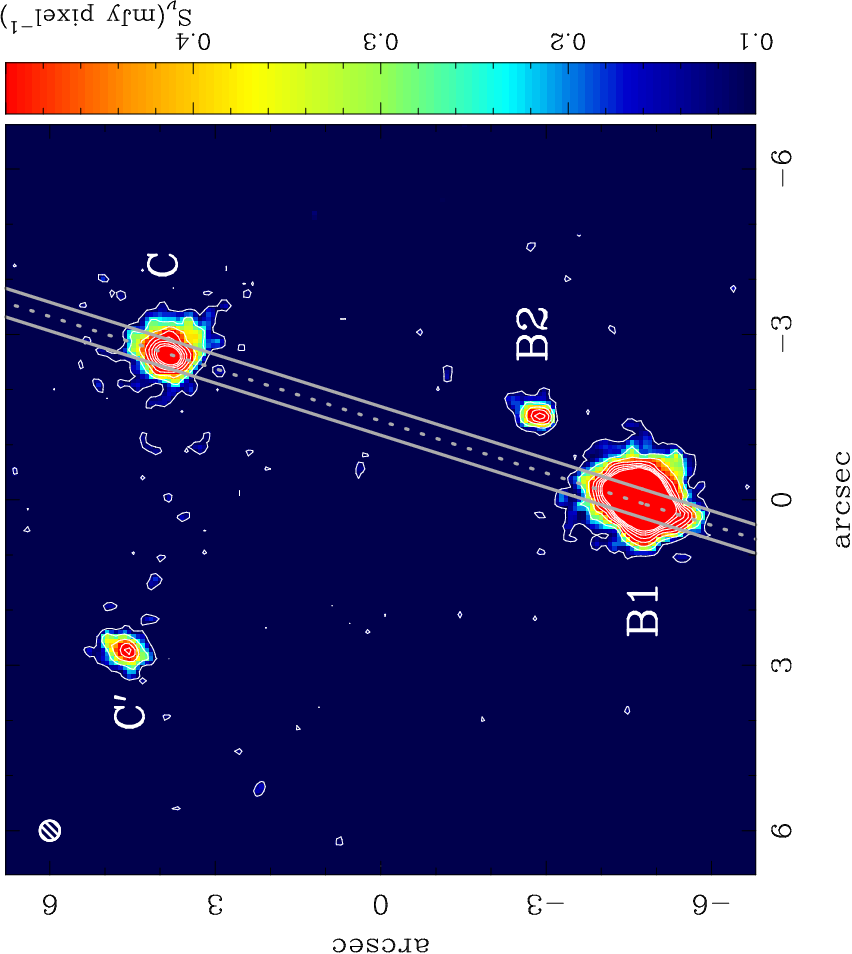}
\caption{GTC/CC  $8.7\,\mu$m acquisition images 
  of IC~694
  (left panel)
  and NGC~3690 (right panel) shown on a
  linear flux scale. 
The projected 
  separation between the galaxy nuclei IC~694/Arp~299-A and NGC~3690/Arp~299-B1
  is approximately 
  21\arcsec \, or 4.5\,kpc, as can be seen from
  figure~5 of AAH09. We also label
  other IR-emitting sources in NGC~3690: B2, C, and C$^\prime$. The lines indicate the
  location, orientation,   and width of 
  the slits. The images have been smoothed with a Gaussian
  function with $\sigma=0.9\,$pixel. The hatched circles at the top left
  of the panels 
  represent the approximate resolution of the images (FWHM). 
Orientation is north up, east to the left. } 
\label{fig:acquisitionimages}
\end{figure*}

The interacting galaxy Arp~299 (Mrk~171, IC~694+NGC~3690) was identified as a 
mid-infrared (MIR) luminous source \citep{Rieke1972} more than 40
years ago. Subsequently, \cite{Gehrz1983} detected 
two bright $10\,\mu$m sources coincident with the galaxy nuclei
\citep[see also][]{Telesco1985}, 
referred to as  Arp~299-A or the nucleus of the eastern
component (IC~694) and Arp~299-B or the nucleus of the western
component (NGC~3690).  Source B is further resolved into B1, the
MIR-bright nucleus, and B2, 
  the optical-bright source (see Fig.~\ref{fig:acquisitionimages}).
Two other bright MIR sources are in the system overlapping region:
Arp~299-C and Arp~299-C$^\prime$.  
The infrared (IR) luminosity of the system is $L_{\rm
  IR}= 6.7 \times 
10^{11}\,L_\odot$ (for a distance $D=44\,$Mpc), which puts it in the
luminous IR galaxy (LIRG) category.

Most of the Arp~299 IR
luminosity arises from intense star formation (SF) activity 
across the two galaxies \citep[see][AAH00 and AAH09, 
hereafter]{Gehrz1983, Charmandaris2002, AAH00, AAH09}. However, there is
also evidence of obscured active galactic
nucleus (AGN) activity in the galaxy nuclei. 

Based on
hard X-ray observations, the
nucleus of NGC~3690/Arp~299-B1 was long suspected to have a Compton-thick
AGN \citep[see e.g.][]{Ballo2004, GonzalezMartin2009,
  Pereira2011}. Moreover, there is a hot dust continuum 
associated with an obscured AGN that contributes substantially to the
nuclear IR emission \citep[][AAH00, AAH09]{Gallais2004}.  
Although initially thought not to be a Seyfert galaxy, 
\cite{GarciaMarin2006} measured Seyfert-like optical line ratios for
Arp~299-B1.

The case for the presence of an AGN in  the nuclear region of 
IC~694/Arp~299-A is not as clear. Apart from the tentative X-ray evidence
\citep{Ballo2004}, perhaps the most convincing argument 
is the identification of a flat spectral radio source among
the large number of compact radio sources detected in the nuclear region
of this galaxy \citep{Neff2004, PerezTorres2010}.

In this paper we present $8-13\,\mu$m spectroscopy of the two nuclei 
of Arp~299 obtained with  
CanariCam \citep[CC, ][]{Telesco2003} on the 10.4\,m Gran Telescopio Canarias
(GTC). The GTC/CC high-angular resolution $0.3-0.6\arcsec$ ($\sim
60-120\,$pc 
for Arp~299 distance)  is approximately a factor of 10 improvement over
 previous MIR spectroscopy of this system
\citep[][AAH09]{Dudley1999,Gallais2004}.  We use the GTC/CC spectroscopy to 
shed light on the processes giving rise to the 
deeply nuclear embedded emission in Arp~299-A and Arp~299-B1.

\section{CanariCam observations  and 
Data Reduction}\label{s:observations}
We obtained mid-IR high-angular resolution 
long-slit spectroscopy  of the nuclear regions of
Arp~299 using CC on the GTC. The observations are  part of our GTC/CC
AGN guaranteed 
time program  (PI C. Telesco). We will use these observations together with
an ESO/GTC large program (PI A. Alonso-Herrero) to conduct a MIR survey of
approximately 100 local AGN. We used the low spectral resolution
$10\,\mu$m grating, which covers the $N$-band $\sim 7.5-13\,\mu$m with
nominal $R=\lambda / \Delta \lambda \sim 175$. A 
0.52\arcsec \, wide slit  was oriented at 345\arcdeg \, so 
that Arp~299-B1 and C could be observed
simultaneously (see
Fig.~\ref{fig:acquisitionimages}, right panel). The plate scale of the
CC $320\times240$  Si:As detector is
0.08\arcsec/pixel, which provides a field of view in imaging mode of
$26\arcsec \times 19\arcsec$. 


The observations were taken in queue mode under photometric conditions
using the  standard MIR chop-nod technique on January 29 and June 25 2013
for Arp~299-A and Arp~299-B1+C, respectively. 
First, we obtained an acquisition image (Fig.~\ref{fig:acquisitionimages})
through the Si-2 filter ($\lambda_{\rm 
  c}=8.7\,\mu$m and $\Delta \lambda_{\rm cut} =1.1\,\mu$m, 50\%
cut-on/off)  to ensure
optimal placement of the slit.
The on-source integration times for the spectroscopy were 1011\,s for
Arp~299-A and  354\,s for  
Arp~299-B1+C.  We also observed  
standard stars in imaging and spectroscopic mode to
provide the photometric calibration, telluric correction, and slit
loss correction. We used the standard stars and galaxy nuclei to
measure an angular resolution at $8.7\,\mu$m.   
of $0.5-0.6\arcsec$ and $0.3\arcsec$ (full width half maximum, FWHM) for  
Arp~299-A and Arp~299-B1+C, respectively.

We reduced the data using the CC 
pipeline {\sc redcan} \citep[][]{GonzalezMartin2013} that includes
stacking of the individual observations, wavelength 
calibration, trace determination, spectral extraction, and flux
calibration. The core emission of the two 
nuclei appears unresolved at the resolution of the GTC/CC acquisition 
images. We  extracted the 1D nuclear spectra in an optimal way for a 
point  source with the aperture size increasing with wavelength to
include all the source flux. Finally, we used the observations of the
standard stars to correct for slit losses. We estimated a 20\% total  
uncertainty of the GTC/CC spectra due to flux calibration and
correction for the point source extraction. 

\begin{figure}

\vspace{0.5cm}

\hspace{0.6cm}
\includegraphics[width=0.3\textwidth,angle=-90]{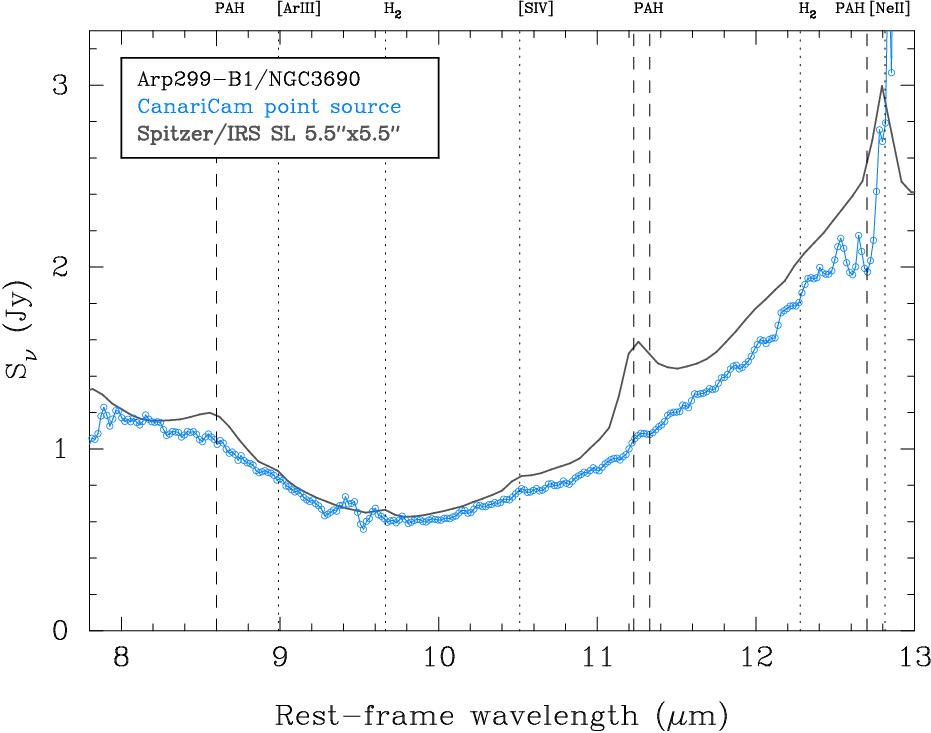}

\smallskip
\hspace{0.6cm}
\includegraphics[width=0.3\textwidth,angle=-90]{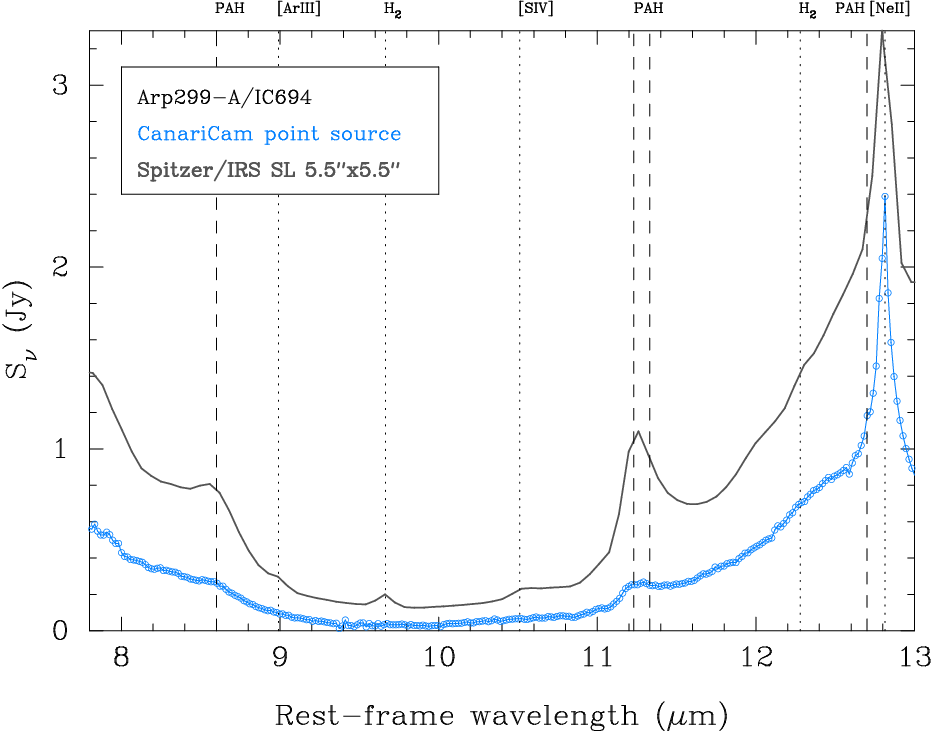}
\caption{GTC/CC spectra (open symbols) and 
  {\it Spitzer}/IRS spectra (black line)
of the nuclear regions of Arp~299-B1 (upper panel) 
and Arp~299-A (lower  panel)   
extracted as point sources. We mark  
emission lines and PAH features. The noise in the GTC/CC spectra
around $9.5\,\mu$m 
is due to the O$_3$ telluric band. }  
\vspace{0.5cm}
\label{fig:spectra}
\end{figure}

To compare with the large scale emission ($\sim 1\,$kpc), we used
spectral mapping 
observations obtained with the Infrared Spectrograph \citep[IRS,
][]{Houck2004} on board {\it Spitzer} (see AAH09 for full details about the
observations). We used {\sc cubism} \citep{Smith2007} 
to extract $3\times3$ pixel ($5.5\arcsec \times 5.5\arcsec$)
Short-Low SL1 ($\sim 7-15\,\mu$m) spectra. We corrected the spectra for point
source emission as  in \cite{Pereira2010}. Fig.~\ref{fig:spectra} shows
the GTC/CC nuclear 
spectra of NGC~3690/Arp~299-B1 (top) and IC~694/Arp~299-A (bottom)
together with the {\it Spitzer}/IRS spectra.

\begin{table}
\center
\caption{{\sc clumpy} torus models for NGC~3690/Arp~299-B1}\label{table:clumpypars}
\begin{tabular}{lccc}
\hline
\hline
{\sc clumpy} torus parameter & Prior/Range & Fit\\
\hline
Torus width, $\sigma_{\rm torus}$           & $15\arcdeg-70\arcdeg$ &
$57\arcdeg^{+7}_{-8}$\\ 
Torus radial thickness, $Y$                 & $5-100$ & $51^{+28}_{-30}$ \\
No. clouds along equator, $N_0$ & $1-15$  & $10^{+3}_{-2}$\\ 
Cloud optical depth, $\tau_V$   & $5-150$ & $45^{+15}_{-13}$\\ 
Cloud radial distribution $r^{-q}$, $q$ & $0-3$   & $2.7^{+0.2}_{-0.3}$\\
Viewing angle, $i$              & $60\arcdeg-90\arcdeg$ & $75\arcdeg^{+8}_{-9}$\\
$A_V({\rm foreground})$        & $16\,{\rm mag}\pm 7$  & $15\,{\rm mag}^{+3}_{-3}$\\
\hline
\end{tabular}

Notes.--- In the {\sc clumpy} models the torus radial thickness is 
defined as $Y=R_{\rm
  o}/R_{\rm d}$, where $R_{\rm o}$ is the outer radius and $R_{\rm d}$
is the   inner radius. 
\smallskip
\end{table}
\smallskip

\section{Nuclear region of NGC~3690/Arp~299-B1}

\subsection{Nuclear versus circumnuclear spectra}
The GTC/CC nuclear spectrum of NGC~3690 was observed under excellent
seeing conditions, ${\rm FWHM}_{8.7\mu{\rm m}}=0.3\arcsec$. Thus the unresolved MIR emission
originates from a region of   $\le 60\,$pc in size. The nuclear
spectrum shows dust continuum 
emission likely 
produced by the obscured AGN and a moderately deep $9.7\,\mu$m
silicate feature. The apparent depth of the 
silicate feature, 
defined as $S_{\rm Si}= \ln ({f_{\rm cont}}/{f_{\rm feature}})$, is
  $S_{\rm Si} = 0.8$. There is 
faint nuclear $11.3\,\mu$m polycyclic aromatic hydrocarbon (PAH)
emission (see Fig.~\ref{fig:spectra}) with an
equivalent width (EW) of $<0.01\,\mu$m.

The $\sim 1\,{\rm
  kpc} \times 1\,{\rm kpc}$ {\it Spitzer}/IRS spectrum shows
 a composite (AGN/SF) nature.  It shows 8.6 and $11.3\,\mu$m PAH
 features together with a strong
continuum due to hot dust \citep[see also][AAH09]{Gallais2004}. 
The EW of the 
$11.3\,\mu$m PAH feature measured using a local continuum is $\sim
0.05\,\mu$m \citep{Pereira2010}, which is typical of AGN
\citep[see][]{HernanCaballero2011}.  
This agrees with the presence of the unresolved IR-bright source B1, 
which has an important hot dust contribution at
$\lambda \ge 2\,\mu$m based on the 
observed near-IR CO index, possibly associated with an AGN. On scales
of a few arcseconds, B1 is
also surrounded by star clusters and bright 
H\,{\sc ii} regions (AAH00).

\subsection{Modeling with the {\sc clumpy} torus models}\label{sec:torusfit}
Clumpy torus models
reproduce satisfactorily the nuclear IR emission of
Seyfert galaxies and give an estimate of the AGN bolometric luminosity
\citep{RamosAlmeida2009,RamosAlmeida2011,Honig2010,AAH11}. In this
section we model the IR spectral energy distribution (SEDs) and
MIR spectrum of NGC~3690  to 
estimate the AGN's bolometric luminosity.

\begin{figure}
\hspace{0.3cm}
\includegraphics[width=0.33\textwidth,angle=-90]{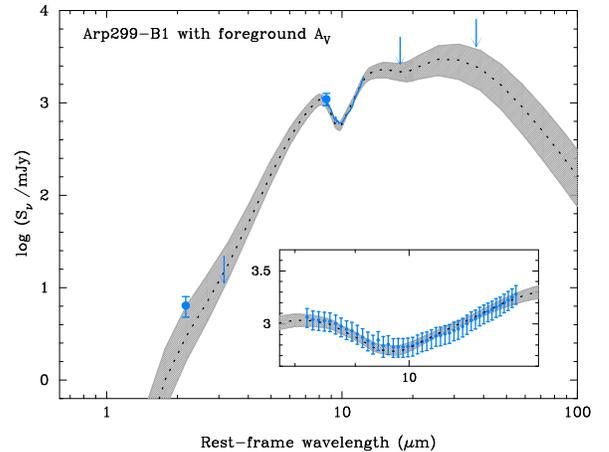}
\caption{Best median fit using the {\sc clumpy} torus models (black
  dotted line) and  
  $1\sigma$ confidence (shaded region) to the nuclear  IR SED (filled blue dots and
  arrows) and MIR 
  GTC/CC 
  spectrum (re-sampled to 45 data points, blue line) of
  NGC~3690/Arp~299-B1. The inset shows the GTC/CC 
  spectrum (filled blue dots with errors) and best fit models as above.}
\label{fig:BCfitNGC3690}
\end{figure}

We constructed the nuclear IR SED of NGC~3690 with 
the  GTC/CC Si-2 $8.7\,\mu$m point source 
measurement of $1050\pm 200$\,mJy 
 and the {\it Hubble Space Telescope}/NICMOS 
$2.2\,\mu$m non-stellar measurement (AAH00). 
We included the Keck $1\arcsec$-diameter
3.2 and $17.9\,\mu$m  \,\citep{Soifer2001} and the
Kuiper Airborne Observatory  
$37.7\,\mu$m 8.5\arcsec-beam flux densities \citep{Charmandaris2002}
as upper 
limit since they are probably contaminated by non-AGN emission. 
Fig.~\ref{fig:BCfitNGC3690} shows the nuclear IR SED.

We used the \cite{Nenkova2008} clumpy torus models, also known as {\sc
  clumpy}, and the Bayesian fitting routine
{\sc bayesclumpy} \citep{AsensioRamos2009} 
to fit the  nuclear IR emission of NGC~3690. A
Bayesian approach  allows to handle properly the
intrinsic degeneracies of the {\sc clumpy}  models, provides probability
distributions of the fitted parameters, and allows the use of
priors for the torus parameters. 

The detection of a nuclear
water maser in Arp~299-B1 \citep{Tarchi2011} implies an almost edge-on
view of the AGN. We
restricted the range of the viewing angles to $i=60\arcdeg-90\arcdeg$ as a
prior. Foreground dust is clearly present in
the nuclear region  of this galaxy (see the optical to
near-IR color map in figure~9 of AAH00). We set the prior for the 
foreground extinction to a Gaussian
distribution centered at 16\,mag with a 7\,mag width, based
on the estimates of  AAH00. We use the extinction 
curve of \cite{Chiar2006}. For the rest of the {\sc clumpy} model parameters we
used the full range (see Table~\ref{table:clumpypars}).

The  best fit {\sc
  clumpy} torus model is shown in Fig.~\ref{fig:BCfitNGC3690} and 
the parameters are listed in Table~\ref{table:clumpypars}. The fit is
achieved with a wide torus, a highly inclined view, and a large number
of clouds along the equatorial direction. This results in a high
covering factor of 
$f_2\sim 0.9$ \citep[see figure~10 in][]{RamosAlmeida2011}. The torus
extinction along the line of sight 
of $A_V^{\rm LOS} ({\rm torus}) \sim 456\,$mag dominates 
over the foreground extinction.  

The best-fit model to the nuclear IR emission provides an
estimate of the AGN bolometric luminosity of $L_{\rm bol}({\rm AGN}) = 3.2
\pm 0.6 \times  10^{44}\,{\rm erg \,s}^{-1}$. This 
is about a factor of 10 higher than the
$0.5-100\,$keV intrinsic luminosity derived for this system with {\it
BeppoSAX} with $N_{\rm H}\sim 1.9 \times 10^{24}\,{\rm cm}^{-2}$
\citep{DellaCeca2002}. Alternatively we can use the 
hard X-ray vs. $12\,\mu$m correlation observed for local Seyfert galaxies
\citep{Gandhi2009, Levenson2009} and the 
nuclear $12\,\mu$m monochromatic luminosity. We predict an intrinsic 
$L_{2-10\,{\rm keV}} = 4.5\pm 1.4 \times 10^{43}\,{\rm erg \, 
  s}^{-1}$. 

The ratio of observed \citep{Ballo2004,Pereira2011} to predicted
intrinsic $2-10\,$keV luminosities is 
of the order of 500, which would only be compatible with the maximum value
of $N_{\rm H}\sim 4 \times 10^{24}\,{\rm cm}^{-2}$ fitted from the {\it
  BeppoSAX} data  \citep{DellaCeca2002}. 
Both MIR estimates of the AGN luminosity are higher than {\it 
  direct} measurements, although the highest possible $N_{\rm H}$
value would produce compatible luminosities.
A  possibility would be that the nuclear MIR emission 
had an important 
contribution from SF. AAH00 estimated that the SF activity in
Arp~299-B1 is approximately five times less than in Arp~299-A. Then using
the nuclear $12\,\mu$m fluxes of Arp~299-A and
Arp~299-B1 we predict that SF  would only contribute $\sim 15\%$ of
the observed $12\,\mu$m nuclear emission from Arp~299-B1. We would
therefore favor the high $N_{\rm H}$ value rather than a strong contamination
from nuclear SF. Finally, it is possible that
the hard X-ray vs. MIR relation might not be applicable to all
Compton-thick AGN. 

\section{Nuclear region of IC~694/Arp~299-A}
\subsection{Nuclear versus circumnuclear spectra}
The GTC/CC nuclear spectrum of IC~694 (Fig.~\ref{fig:spectra}) shows
clear PAH emission at 8.6 and $11.3\,\mu$m indicative of the presence
of SF on nuclear scales of $\le 120\,$pc. This is consistent with
results from high angular resolution near-IR (AAH00) and radio
\citep{PerezTorres2009} observations of this nucleus. 
We measured a GTC/CC nuclear EW($11.3\,\mu{\rm m \, PAH})=0.12\pm
0.01\,\mu$m, which is 
lower than that measured from the kpc-scale {\it Spitzer}/IRS
spectrum  \citep[$0.23\pm0.01\,\mu$m, see][]{Pereira2010}. 
These are typical of galaxies with a composite (AGN/SF) activity 
\citep{HernanCaballero2011}. Moreover, the 
decreased nuclear  EW with respect to the 1\,kpc-scale one 
also indicates a higher relative 
contribution in the nuclear region from  continuum emission, 
possibly produced by dust 
heated by the putative AGN. This is similar to findings for nearby
Seyferts with nuclear SF activity
\citep{Esquej2013}.

\subsection{Fit to the nuclear spectrum}
\begin{figure}
\hspace{0.3cm}
\includegraphics[width=0.35\textwidth,angle=-90]{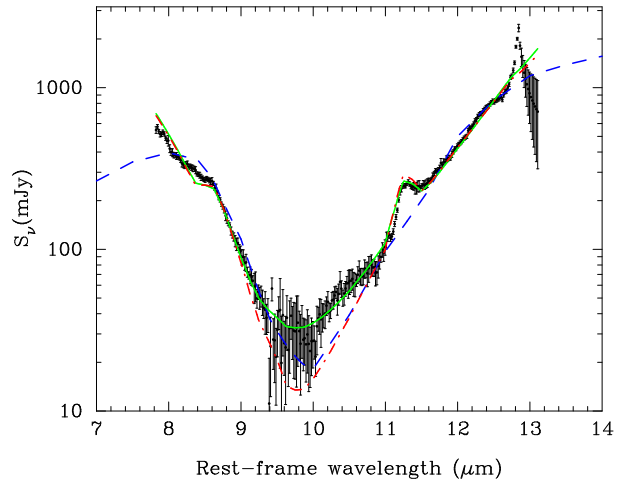}
\caption{Fit to the GTC/CC nuclear spectrum of IC~694/Arp~299-A. The error bars
  used for the fit 
are estimated from the scatter of the points in narrow wavelength
bins. The blue dashed line is the best fit with the spherical dusty
shell models of   \cite{Levenson2007}. The dot-dashed red line is the
  fit using a   black-body + the Orion bar PAH spectrum with silicate
  absorption 
  as modelled with the Trapezium silicate grains following
    \cite{Roche2007}. The solid green line 
  includes additionally a power-law component that fills  in the $\sim
  9-11\,\mu$m   spectral range. The error bars around the [Ne\,{\sc
      ii}]$12.81\,\mu$m line 
  were increased artificially to exclude it from the fit.} 
\vspace{0.1cm}
\label{fig:dustfitIC694}
\end{figure}

Deep silicate features, as in
the nuclear spectrum of Arp 299-A, cannot be reproduced with the {\sc 
  clumpy} torus models \citep{Levenson2007, AAH11,
  GonzalezMartin2013}.  Instead,  \cite{Levenson2007} 
  demonstrated that spherical dusty shell models where the nuclear 
source is deeply embedded  in a smooth distribution of geometrically
and optically  thick material can produce such deep
features. Moreover, the reprocessed emission in these models does not
depend on the input spectrum of the heating source. We
obtained the best fit
shell model to the GTC/CC spectrum of Arp299-A with  a total optical
depth through the shell of $\tau_V = 
162$, a maximum temperature at the inside
surface of the shell of $T_{\rm max}=1500\,$K, $Y=100$ (defined as in
Table~\ref{table:clumpypars}), and an $r^{-1}$ density profile. These
are similar to the values obtained for the deep silicate feature
observed in the ultraluminous IR galaxy 
IRAS~08572+3915 \citep{Levenson2007}. While
the overall fit to Arp~299-A is reasonable, it fails to reproduce the shape of
the silicate feature around $10-11\,\mu$m (see
Fig.~\ref{fig:dustfitIC694}). Also, the dusty 
shell model
does not fit the strong nuclear PAH emission,  
as it does not include a star formation component.

Alternatively, we fitted the GTC/CC spectrum with a
combination of emission and absorption components using the technique
described by \citet[][]{Roche2007}. We first tried a black-body  + the
Orion Bar PAH 
spectrum suffering silicate absorption. This fit is shown as a dashed line
in Fig.~\ref{fig:dustfitIC694} and has $\chi^2/N=5.9$, where $N$ is
the number of degrees of freedom. The silicate optical
depth is $\tau_{9.7\mu{\rm m}} = 1.4$ ($A_V=21\,$mag,  assuming a
  dust screen geometry)
with the Trapezium silicate grain profile, which is representative of
those of molecular clouds. Using the curve of the $\mu$ Cephei supergiant
gave a poorer fit, but these silicate grains seem to be more appropriate for
sources with extremely deep silicate features \citep{Roche2007}.  

Although these two components provide quite a good qualitative
agreement over most of the spectrum,  
the fit falls below the data points between 9 and $11\,\mu$m, as
 found with the dusty shell model fit above.
To improve the fit, we added an IR power-law component \citep[$f_\nu \propto
  \nu^{-2}$, see][]{RamosAlmeida2009}.
This fills in the $9-11\,\mu$m spectrum (Fig.~\ref{fig:dustfitIC694}) and
improves the fit ($\chi^2/N=4.1$). The derived
silicate optical depth is 
$\tau_{9.7\mu{\rm m}} = 1.6$ or $A_V=24\,$mag, in
agreement with other IR estimates (Gallais et al. 2004, AAH00, AAH09).

The power-law
component contributes approximately 35\% of the observed $9\,\mu$m
flux density in the nuclear GTC/CC spectrum of IC~694. We propose that
this emitting component might be associated with dust heated 
by the putative AGN  and also be responsible for the decreased nuclear
EW($11.3\,\mu$m PAH). 
From the power-law monochromatic $12\,\mu$m luminosity we estimated an
intrinsic  
$L_{2-10{\rm keV}}\sim 1 \pm 0.2 \times 10^{43}\,{\rm
  erg \,s}^{-1}$ using  the \cite{Gandhi2009} relation, as in
Section~\ref{sec:torusfit}. The putative AGN in IC~694
would be about $5$ times less luminous than that in NGC~3690.

\section{Discussion and conclusions}
We obtained GTC/CC
high-angular resolution ($0.3-0.6\arcsec$, $\sim 60-120$\,pc) MIR
spectra of the interacting galaxy Arp~299. We presented evidence 
of deeply embedded AGN   activity in both nuclei thanks
to the improved angular resolution  of these data of almost a factor
of 10 with respect to previous MIR spectroscopy of this
system. 

The GTC/CC nuclear spectrum of NGC~3690/Arp~299-B1 shows a strong dust
continuum with a moderate  silicate feature in absorption. 
The nuclear $11.3\,\mu$m PAH feature is weak compared with that
in the 1\,kpc-scale {\it Spitzer}/IRS spectrum. We fitted the
IR emission of NGC~3690 with the {\sc clumpy} models. The dust is
located 
in a  high covering factor torus and heated 
by an AGN with $L_{\rm bol}({\rm AGN})=3.2
\pm 0.6 \times 10^{44}\,{\rm erg \,s}^{-1}$. The torus
high extinction along the line of sight 
dominates over the foreground extinction. 

The GTC/CC nuclear (region $\le 120\,$pc in size) 
spectrum of IC~694/Arp~299-A shows a very
deep silicate feature reflecting the highly embedded nature of this
source with a fitted $\tau_{9.7\mu{\rm m}}=1.6$ ($A_V=24\,$mag, 
  for a dust screen geometry). 
There is strong nuclear SF based on the detection of the $8.6\,\mu$m and
$11.3\,\mu$m PAH features. The lower EW of the nuclear $11.3\,\mu$m
PAH feature compared with the
kpc-scale IRS value suggests the presence of a MIR continuum 
that also fills in the silicate feature within $\sim
9-11\,\mu$m. If this nuclear continuum emitting 
component is due to dust heated by an AGN, then we
estimated an AGN luminosity  about five times less  than that in NGC~3690.

The interaction of the  Arp~299 system started at least
750\,Myr ago \citep{Hibbard1999} but the system has not yet 
fully merged. The projected nuclear separation is $\sim 4.5\,$kpc
and the system is experiencing intense SF spread
out over several kpc (Charmandaris et al. 2002, AAH09). Our work
provided evidence of dual  dust enshrouded  
AGN, probably indicative of an early phase of AGN activity. 
The combined luminosities of the two AGN only account for
$\sim 15\%$ of the total IR luminosity of the system. This relatively
small AGN contribution is typical of local LIRGs
\citep[see][]{Petric2011,AAH12}. 

Numerical simulations predict that interaction-induced AGN 
activity is common. However, dual AGN activity with $L_{\rm bol}({\rm AGN}) >
10^{44}\,{\rm erg \,s}^{-1}$ would only occur simultaneously
for approximately 10\% of the duration of the interaction \citep[see
e.g.,][]{vanWassenhove2012}. Therefore,
Arp~299 represents an interesting case study to test theoretical
predictions for dual AGN activity during the early stages of galaxy
interactions.

\smallskip
We are extremely grateful to the GTC staff for their constant and
enthusiastic support. We also thank an anynomous referee for
  comments that helped improve the paper. The following Spanish Plan Nacional 
de Astronom\'{\i}a y Astrof\'{\i}sica  grants are acknowledged: 
AYA2009-05705-E (AAH, PE, CRA, and MPS),  AYA2010-21887-C04 (CRA and JMRE), 
AYA2010-18029 (AAR), AYA2010-21161-C02-01 (LC), and
AYA2012-39168-C03-01 (JMRE and OGM). AAR also acknowledges financial support 
through the Ram\'on y Cajal fellowships and Consolider-Ingenio 2010
CSD2009-00038. CP and CMT acknowledge support 
from NSF grants 0904421 and AST-903672, respectively.
NAL and RM are supported by the Gemini Observatory, which is
operated by the Association of Universities
for Research in Astronomy, Inc., on behalf of the international Gemini
partnership of Argentina,
Australia, Brazil, Canada, Chile, and the United States of America.

Based on observations made with the Gran Telescopio Canarias (GTC),
installed in the Spanish Observatorio del Roque de los Muchachos of the
Instituto de Astrof\'{\i}sica de Canarias, in the island of La Palma.
Based party on observations obtained with
the {\it Spitzer Space Observatory}, which is operated by JPL,
Caltech, under NASA contract 1407.
This research has made use of the NASA/IPAC Extragalactic Database
(NED) which is operated by JPL, Caltech, 
under contract with the National Aeronautics
and Space Administration.

\end{document}